\documentclass[12pt]{amsart}
\usepackage{amsmath,amssymb,hyperref,enumitem,graphicx}
\usepackage[sort&compress,numbers]{natbib}

\hypersetup{colorlinks=true,allcolors=blue}
\usepackage[all]{hypcap}

\usepackage[foot]{amsaddr}
\usepackage[sc]{mathpazo}

\newcommand\ack{\subsection*{Acknowledgment}}

\DeclareMathAlphabet\mathsfbi{T1}{phv}{b}{it}

\numberwithin{equation}{section}

\newcommand\BV{\boldsymbol} 
\newcommand\BM{\mathsfbi} 

\newcommand\dif{\:\!\mathrm{d}}
\newcommand\deriv[2]{\frac{\mathrm{d} #1}{\mathrm{d} #2}}
\newcommand\parderiv[2]{\frac{\partial #1}{\partial #2}}

\newcommand\trace{\mathrm{tr}}

\newcommand\tu{\tilde{\BV u}}
\newcommand\tP{\,\tilde{\!\BM P}}
\newcommand\tS{\,\tilde{\!\BM S}}
\newcommand\tq{\tilde{\BV q}}
\newcommand\tQ{\,\tilde{\!\BM Q}}
\newcommand\tomega{\tilde{\BV\omega}}

\newcommand\myatop[2]{\genfrac{}{}{0pt}{}{#1}{#2}}

\def \Rey {\textit{Re}}

\begin{document}

\author[Rafail V. Abramov]{Rafail V. Abramov}

\address{Department of Mathematics, Statistics and Computer Science,
University of Illinois at Chicago, 851 S. Morgan st., Chicago, IL 60607}

\email{abramov@uic.edu}

\title{Turbulence in large scale two-dimensional balanced hard sphere
  gas flow}

\begin{abstract}
In recent works we developed a model of balanced gas flow where the
momentum equation possesses an additional mean field forcing term,
which originates from the hard sphere interaction potential between
the gas particles. We demonstrated that, in our model, a turbulent gas
flow with a Kolmogorov kinetic energy spectrum develops from an
otherwise laminar initial jet. In the current work, we investigate the
possibility of a similar turbulent flow developing in a large scale
two-dimensional setting, where a strong external acceleration
compresses the gas into a relatively thin slab along the third
dimension. The main motivation behind the current work is the
following. According to observations, horizontal turbulent motions in
the Earth atmosphere manifest in a wide range of spatial scales, from
hundreds of meters to thousands of kilometers. Yet, the air density
rapidly decays with altitude, roughly by an order of magnitude each
15-20 kilometers. This naturally raises the question as to whether or
not there exists a dynamical mechanism which can produce large scale
turbulence within a purely two-dimensional gas flow. To our surprise,
we discover that our model indeed produces turbulent flows and the
corresponding Kolmogorov energy spectra in such a two-dimensional
setting.
\end{abstract}

\maketitle

\section{Introduction}

In his famous work, \citet{Rey83} demonstrated that an initially
laminar flow of a liquid consistently develops turbulent motions
whenever the high Reynolds number condition is satisfied. Later,
\citet{Kol41a,Kol41b,Kol41c} and \citet{Obu41,Obu49,Obu62} observed
that the time-averaged Fourier spectra of the kinetic energy of an
atmospheric flow possess a universal decay structure, corresponding to
the inverse five-thirds power of the Fourier wavenumber. With help of
detailed measurements from the Global Atmospheric Sampling Program,
\citet{NasGag} estimated the energy spectra of meridional and zonal
atmospheric winds, and found that the spectra exhibited the inverse
cubic power scaling at large scales, and the inverse five-thirds power
scaling at moderate and small scales. As of current, the physics of
turbulence formation in a laminar flow, as well as the origin of power
scaling of turbulent kinetic energy spectra, remain unknown.

In our recent work \citep{Abr22}, we considered a system of many
particles, each of mass $m$, interacting solely via a repelling
short-range potential $\phi(r)$, with the convention that $\phi(r)\to
0$ as $r\to\infty$. In the limit of infinitely many such particles, we
obtained, via the standard Bogoliubov--Born--Green--Kirkwood--Yvon
(BBGKY) formalism~\citep{Bog,BorGre,Kir}, the following Vlasov-type
equation~\citep{Vla} for the mass-weighted distribution density of a
single particle $f(t,\BV x,\BV v)$:
\begin{equation}
\label{eq:vlasov}
\parderiv ft+\BV v\cdot\parderiv f{\BV x}=\frac 1\rho\parderiv{
  \bar\phi}{\BV x}\cdot\parderiv f{\BV v}.
\end{equation}
Above, the mean field potential $\bar\phi$ is given via
\begin{equation}
\label{eq:bphi}
\bar\phi(t,\BV x)=\frac 1{2m}\rho(t,\BV x)p(t,\BV x)\int\left[ 1-\exp
  \left(-\frac{\phi(\|\BV y\|)}{\theta(t,\BV x)}\right)\right] \dif\BV
y.
\end{equation}
The quantities $\rho$, $p$ and $\theta$ are the mass density, pressure
and kinetic temperature of $f$, respectively, given, together with the
mean velocity $\BV u$, via
\begin{equation}
\label{eq:rho_u_p}
\rho=\int f\dif\BV v,\qquad\rho\BV u=\int\BV vf\dif\BV v,\qquad p=
\rho\theta=\frac 13\int\|\BV v-\BV u\|^2f\dif\BV v.
\end{equation}
Further, computing the usual hierarchy of the transport equations for
the velocity moments of $f$, and closing it under the infinite
Reynolds number assumption, in \citep{Abr22} we arrived at the
following equations for a balanced compressible gas flow:
\begin{equation}
\label{eq:bflow3D}
\parderiv\rho t+\nabla\cdot(\rho\BV u)=0,\qquad\parderiv{(\rho\BV u)}t
+\nabla\cdot(\rho\BV u^2)+\nabla p=-\nabla\bar\phi,\qquad\parderiv pt+
\BV u\cdot\nabla p=0.
\end{equation}
The difference between the equations above and the conventional
compressible Euler equations \citep{Bat,Gols} is that the former
preserve the pressure $p$ along the stream lines (balanced flow), and
have the mean field forcing $-\nabla\bar\phi$ in the momentum equation.

In \cite{Abr22}, we numerically simulated the equations in
\eqref{eq:bflow3D} for a gas of hard spheres with the mass and
diameter corresponding to those of argon, flowing through a straight
three-dimensional pipe. For the initial condition of that simulation,
we selected a straight laminar Bernoulli jet, which is a steady state
in the conventional Euler equations. We observed, however, that in our
model \eqref{eq:bflow3D} such a jet quickly develops into a fully
turbulent flow. We also examined the Fourier spectrum of the kinetic
energy of the simulated flow, and found that its time average decayed
with the rate of inverse five-third power of the wavenumber, which
corresponded to the famous Kolmogorov spectrum.

The main goal of the current work is to develop and test a similar
framework for a gas flow under a strong gravity acceleration, which,
on a large scale, is effectively two-dimensional, and, therefore, it
may be impractical to introduce the vertical direction from the
computational perspective. While it is obvious that such a
two-dimensional flow, by its simplified nature, lacks many features of
a full three-dimensional flow (even with the latter confined to a
relatively thin slab), our goal here is to examine whether or not
turbulent features could manifest in such a simplified two-dimensional
model, which is severely restricted in other physical aspects by its
own low dimensionality. This is important, in particular, for
long-range climate prediction models, where the restricted dimension
may be necessitated by the requirement of a long-range simulation.

The work is organized as follows. In Section~\ref{sec:vlasov} we
derive the Vlasov equation which incorporates an external acceleration
in a potential form. In Section~\ref{sec:moment_equations} we assume
that the strong constant external acceleration is applied in the
vertical direction, and further arrive at the system of equations for
a balanced flow, confined to a horizontal plane. In
Section~\ref{sec:numerical} we show the results of numerical
simulations of a large scale flow in two scenarios -- one is an
inertial jet, and another is a cyclostrophic vortex. We observe that
turbulent motions with Kolmogorov spectra manifest in both scenarios
from laminar initial conditions. Section~\ref{sec:summary} summarizes
the results of the work.

\section{The Vlasov-type equation with external acceleration}
\label{sec:vlasov}

Here we repeat the derivation of the Vlasov-type equation in
\eqref{eq:vlasov} for the motion of particles under an external
acceleration. In the presence of an external acceleration, the
equations of motion for $N$ particles interacting via a potential
$\phi(r)$ are given via
\begin{equation}
\deriv{\BV x_i}t=\BV v_i,\qquad \deriv{\BV v_i}t=\BV g(\BV x_i)-
\parderiv{ }{\BV x_i}\sum_{\myatop{j=1}{j\neq i}}^N\phi(\|\BV x_i-\BV
x_j\|),
\end{equation}
where $\BV g(\BV x)$ is the external acceleration which acts on any
particle at the location $\BV x$, and depends only on the coordinates
of that particle. Further, we will assume that $\BV g(\BV x)$ is a
potential acceleration, that is, there exists a potential function
$h(\BV x)$ such that $\BV g(\BV x)=-\partial h(\BV x)/\partial\BV
x$. As in our previous works \citep{Abr21,Abr22}, we concatenate
\begin{equation}
\BV X=(\BV x_1,\ldots,\BV x_N),\qquad\BV V=(\BV v_1,\ldots,\BV v_N),
\qquad\Phi(\BV X)=\sum_{i=1}^{N-1}\sum_{j=i+1}^N\phi(\|\BV x_i-\BV x_j
\|).
\end{equation}
Additionally, due to the presence of the external acceleration, here
we concatenate
\begin{equation}
H(\BV X)=\sum_{i=1}^N h(\BV x_i).
\end{equation}
In the new notations, the Liouville equation for the density of states
$F(t,\BV X,\BV V)$ is given via
\begin{equation}
\label{eq:liouville}
\parderiv Ft+\BV V\cdot\parderiv F{\BV X}=\parderiv{(H+\Phi)}{\BV X}
\cdot\parderiv F{\BV V}.
\end{equation}
Due to the presence of the external acceleration, the total momentum
of the system is no longer a first integral. However, due to the fact
that $\BV g(\BV x)$ is a potential acceleration, the total energy $E$
of the system remains an invariant of motion:
\begin{equation}
E=\frac{\|\BV V\|^2}2+H(\BV X)+\Phi(\BV X)=\text{const}.
\end{equation}
Therefore, similarly to \citep{Abr21,Abr22}, any suitable function of
the form
\begin{equation}
F_0(\BV X,\BV V)=F_0(E)=F_0\big(\|\BV V\|^2/2+H(\BV X)+\Phi(\BV
X)\big)
\end{equation}
is automatically a steady state of the Liouville equation in
\eqref{eq:liouville}. Among those functions, the canonical Gibbs
equilibrium state is given via
\begin{equation}
F_G=\frac 1{(2\pi\theta_0)^{3N/2}Z_N}\exp\left(-\frac{\|\BV V\|^2/2
  +H+\Phi }{\theta_0}\right),\quad Z_N=\int e^{-(H+\Phi)/\theta_0}
\dif\BV X,
\end{equation}
where $\theta_0$ is the equilibrium kinetic temperature of the system.
For the purpose of the work, below we will assume that the gas is {\em
  dilute} (that is, the mean free path of a particle is much larger
than the range of the potential interaction), and thus the
contribution of $\Phi(\BV X)$ to the integral for $Z_N$ is negligible:
\begin{equation}
Z_N=\int e^{-(H+\Phi)/\theta_0}\dif\BV X\approx\int e^{-H/\theta_0}
\dif\BV X=\left(\int e^{-h/\theta_0}\dif\BV x\right)^N =Z_1^N.
\end{equation}
The latter leads to the following form of the canonical Gibbs state
for a dilute gas:
\begin{equation}
\label{eq:F_G}
F_G(\BV X,\BV V)=e^{-\Phi(\BV X)/\theta_0}\prod_{i=1}^N f_G(\BV x_i,
\BV v_i),\qquad f_G(\BV x,\BV v)=\frac {e^{-(\|\BV v\|^2/2+h(\BV
    x))/\theta_0}}{(2\pi\theta_0)^{3/2} Z_1}.
\end{equation}

\subsection{Preservation of the R\'enyi divergences}

Despite the presence of the external acceleration, it can be easily
shown that the Liouville equation in \eqref{eq:liouville} still
preserves the family of general R\'enyi divergences \citep{Ren}, as
long as the integration by parts is not affected by the boundary
conditions. Indeed, for two differentiable functions $\psi_1$ and
$\psi_2$, we have
\begin{multline}
\parderiv{}t\int\psi_1(F)\psi_2(F_0)\dif\BV X\dif\BV V=\int\parderiv{
  \psi_1(F)}t\psi_2(F_0)\dif\BV X\dif\BV V=\int\psi_2(F_0) \bigg(
\parderiv{(H+\Phi)}{\BV X}\cdot\parderiv{}{\BV V}-\\-\BV V\cdot
\parderiv{}{\BV X} \bigg)\psi_1(F)\dif\BV X\dif\BV V=\int\psi_1(F)
\bigg(\BV V\cdot \parderiv{}{\BV X}- \parderiv{(H+\Phi)}{\BV X}\cdot
\parderiv{}{\BV V} \bigg)\psi_2(F_0)\dif\BV X\dif\BV V=0.
\end{multline}
Choosing $\psi_1(x)=x^\alpha$, $\psi_2(x)=x^{1-\alpha}$ for some
parameter $\alpha>0$ leads to the preservation of the R\'enyi
divergence $D_\alpha(F,F_0)$:
\begin{equation}
D_\alpha(F,F_0)=\frac 1{\alpha-1}\ln\int F^\alpha F_0^{1-\alpha}
\dif\BV X\dif\BV V.
\end{equation}
The Kullback--Leibler divergence \citep{KulLei} is a special case of
the R\'enyi divergence with $\alpha=1$. As we noted in
\cite{Abr21,Abr22}, the preservation of the R\'enyi divergences plays
a similar role to Boltzmann's $H$-theorem -- that is, if the initial
state of $F$ is close to $F_0$ in the sense of any R\'enyi metric,
then the solution will also remain a nearby state.

\subsection{Joint two-particle marginal distribution of the Gibbs
state}

In what follows, it is important to note the structure of the joint
two-particle marginal distribution $F^{(2)}_G$ of the Gibbs state for
a dilute gas in \eqref{eq:F_G}. Integrating $F_G$ in \eqref{eq:F_G}
over all particles but the first two, and taking advantage of the
dilute gas assumption, one easily obtains
\begin{equation}
\label{eq:F^2_G}
F^{(2)}_G(\BV x_1,\BV v_1,\BV x_2,\BV v_2)=\int F_G\dif\BV x_3\dif\BV
v_3\ldots\dif\BV x_N\dif\BV v_N=f_G(\BV x_1,\BV v_1)f_G(\BV x_2,\BV
v_2)e^{-\phi(\|\BV x_2-\BV x_1\|)/\theta_0},
\end{equation}
where $f_G(\BV x,\BV v)$ is the Gibbs state for a single particle.
Here, note that, even at a thermodynamic equilibrium, the particles in
a pair cannot be regarded as being independent, since there is a
spatial correlation term through the interaction potential $\phi$.

\subsection{Transport equation for a single particle}

We denote the marginal distribution for the first particle via $f$,
that is,
\begin{equation}
f(t,\BV x,\BV v)=\int F\dif\BV x_2\dif\BV v_2\ldots\dif\BV
x_N\dif\BV v_N,
\end{equation}
where, for convenience, we drop the subscripts from $\BV x_1$ and $\BV
v_1$. We then integrate the Liouville equation in \eqref{eq:liouville}
over $\dif\BV x_2\dif\BV v_2 \ldots\dif\BV x_N\dif\BV v_N$:
\begin{equation}
\parderiv ft+\BV v\cdot\parderiv f{\BV x}+\BV g\cdot\parderiv f{\BV
  v}=\sum_{i=2}^N\int\left(\parderiv{}{\BV x}\phi(\|\BV x-\BV y\|)
\cdot\parderiv{}{\BV v}-\BV w\cdot\parderiv{}{\BV y}\right)F^{(2)
}_{1,i}(\BV x,\BV v,\BV y,\BV w)\dif\BV y\dif\BV w,
\end{equation}
where $F^{(2)}_{1,i}$ is the joint distribution of the first and
$i$-th particles, that is,
\begin{equation}
F^{(2)}_{1,i}(\BV x,\BV v,\BV x_i,\BV v_i)=\int F\dif\BV x_2\dif\BV
v_2\ldots\dif\BV x_{i-1}\dif\BV v_{i-1}\dif\BV x_{i+1}\dif\BV
v_{i+1}\ldots\dif\BV x_N\dif\BV v_N,
\end{equation}
and $\BV y$, $\BV w$ serve as dummy variables of integration for the
coordinate and velocity, respectively, of the $i$-th particle.
Observe that there cannot be any boundary effects in the velocity
integration, because, realistically, $F$ and its derivatives must
vanish at infinite velocities. However, there could still be boundary
effects in the spatial integration, since the spatial domain is
generally bounded, and the presence of the external acceleration $\BV
g$ introduces spatial anisotropy. Thus, we retain the terms with
spatial derivatives of the joint two-particle distribution in the
equation for now.

\subsection{A closure for the single particle transport equation}

To obtain the transport equation for $f$ alone, we need to express
$F^{(2)}_{1,i}$ via $f$. For that, a standard assumption is that all
pairs of particles are identically distributed \citep{Abr17,Abr22}, so
that the transport equation becomes
\begin{equation}
\parderiv ft+\BV v\cdot\parderiv f{\BV x}+\BV g\cdot\parderiv f{\BV
  v}=(N-1)\int\left(\parderiv{}{\BV x}\phi(\|\BV x-\BV y\|) \cdot
\parderiv{}{\BV v}-\BV w\cdot\parderiv{}{\BV y}\right)F^{(2)}(\BV x,
\BV v,\BV y,\BV w)\dif\BV y\dif\BV w.
\end{equation}
Following our recent work \citep{Abr22}, we assume that $F^{(2)}$ has
the same form as its corresponding equilibrium Gibbs distribution,
that is
\begin{equation}
F^{(2)}(\BV x,\BV v,\BV y,\BV w)=f(\BV x,\BV v)f(\BV y,\BV w)\exp
\bigg(-\frac{\phi(\|\BV x-\BV y\|)}{\theta(\frac{\BV x+\BV
    y}2)}\bigg).
\end{equation}
Here, the kinetic temperature $\theta$ can no longer be regarded as
being at equilibrium, and, instead, we take it at the midpoint between
the two particles, for symmetry.

Next, we renormalize $f$ so that it becomes the mass density, that is,
$f\to Nmf$. This leads, as $N\to\infty$, to
\begin{equation}
\parderiv ft+\BV v\cdot\parderiv f{\BV x}=\left(\frac 1\rho\parderiv{
  \bar\phi}{\BV x}-\BV g\right)\cdot\parderiv f{\BV v}-\frac fm\int
\parderiv{}{\BV y}\cdot\bigg[\rho(\BV y)\BV u(\BV y)\exp\bigg(-\frac{
    \phi(\|\BV x-\BV y\|)}{\theta(\frac{\BV x+\BV y}2)}\bigg)\bigg]
\dif\BV y,
\end{equation}
where we integrated the product $\BV wf(\BV w)$ over $\dif\BV w$ and
obtained $\rho\BV u$, according to \eqref{eq:rho_u_p}. Via the
divergence theorem, the volume integral in the right-hand side can be
expressed as the surface integral over the domain boundary $S$,
\begin{multline}
\int\parderiv{}{\BV y}\cdot\bigg[\rho(\BV y)\BV u(\BV y)\exp\bigg(
  -\frac{\phi(\|\BV x-\BV y\|)}{\theta(\frac{\BV x+\BV y}2)}\bigg)
  \bigg]\dif\BV y=\\=\int(\rho\BV u)|_S\exp\bigg(-\frac{\phi(\|\BV
  x-\BV y_S\|)}{\theta(\frac{\BV x +\BV y_S}2)}\bigg)\cdot\BV n\dif
S=\int(\rho\BV u)|_S\cdot\BV n\dif S=-[\rho\BV u]_{\text{in}},
\end{multline}
where $[\rho\BV u]_{\text{in}}$ is the net {\em inward} momentum flux
through the boundary. Above we note that $\BV x$ is located inside the
domain, while $\BV y_S$ is on the boundary, which allows us to set the
exponential to 1 as long as the range of potential is sufficiently
short. This leads to
\begin{equation}
\parderiv ft+\BV v\cdot\parderiv f{\BV x}=\left(\frac 1\rho\parderiv{
  \bar\phi}{\BV x}-\BV g\right)\cdot\parderiv f{\BV v}+\frac fm[\rho
  \BV u]_{\text{in}}.
\end{equation}
In a practical scenario, we can assume that the net momentum flux
through the boundary is zero; indeed, if we have a horizontal slab
domain with strong gravity and impenetrable bottom, the momentum
fluxes through the top and bottom are zero (the former due to
exponentially vanishing density, the latter due to impenetrability),
and we can additionally assume that the net momentum flux through the
sides is also zero if the total amount of the gas particles inside our
domain remains fixed. This further leads to the following Vlasov-type
equation for $f$ with an external acceleration:
\begin{equation}
\label{eq:vlasov_g}
\parderiv ft+\BV v\cdot\parderiv f{\BV x}=\left(\frac 1\rho\parderiv{
  \bar\phi}{\BV x}-\BV g\right)\cdot\parderiv f{\BV v}.
\end{equation}

\section{The two-dimensional moment equations in the presence of
  gravity}
\label{sec:moment_equations}

Here, we are going to assume that the vector of external acceleration
$\BV g$ is constant, and that the reference frame is chosen so that
$\BV g$ points in the opposite direction of the $z$-axis (that is,
$\BV g=(0,0,-g)$, and, as a consequence, $h(\BV x)=gz$). In this
situation, it is convenient to separate the variables into the
horizontal and vertical ones; from now on, $\BV x=(x,y)$ and $\BV
v=(v_x,v_y)$ will denote the horizontal coordinate and velocity
vectors, respectively, while $z$ and $w$ will refer to the vertical
coordinate and velocity, respectively. Accordingly,
$\nabla=(\partial/\partial x,\partial/\partial y)$ will refer to the
horizontal spatial differentiation. In the new notations, the
gravity-forced Vlasov-type equation in \eqref{eq:vlasov_g} becomes
\begin{equation}
\parderiv ft+\BV v\cdot\nabla f+w\parderiv fz=\frac{\nabla\bar\phi}
\rho\cdot\parderiv f{\BV v}+\left(g+\frac 1\rho\parderiv{\bar\phi}
z\right)\parderiv fw.
\end{equation}
We now assume that the horizontal dimensions of the domain are much
larger than the characteristic scale of the exponential decay of $f$
in the vertical dimension. As a result, it can be assumed that, on the
horizontal scale, the dynamics are largely two-dimensional, with the
vertical dimension supplying certain parameterized effects. To
describe this, we now integrate $f$ over $\dif z$:
\begin{equation}
\tilde f(t,\BV x,\BV v,w)=\int_0^\infty f(t,\BV x,z,\BV v,w)\dif z.
\end{equation}
which results in the advection term of the form
\begin{equation}
\int_0^\infty\left(\parderiv ft+\BV v\cdot\nabla f+w\parderiv fz
\right)\dif z= \parderiv{\tilde f}t+\BV v\cdot\nabla\tilde f-w f
|_{z=0},
\end{equation}
as $f$ vanishes exponentially with altitude. This leads to the
following Vlasov-type equation for $\tilde f$:
\begin{equation}
\label{eq:vlasov_g_tf}
\parderiv{\tilde f}t+\BV v\cdot\nabla\tilde f=wf|_{z=0}+\parderiv{}{
  \BV v}\cdot\int_0^\infty\frac{\nabla\bar\phi}\rho f\dif z+\parderiv{
}w\int_0^\infty\left(g+\frac 1\rho\parderiv{\bar\phi}z\right)f\dif z.
\end{equation}

\subsection{The velocity moment equations}

For the zero-order moment we integrate the Vlasov-type equation for
$\tilde f$ in \eqref{eq:vlasov_g_tf} over $\dif\BV v\dif w$, with the
notations
\begin{equation}
\tilde\rho=\int\tilde f\dif\BV v\dif w,\qquad\tilde\rho\tu=\int\BV
v\tilde f\dif\BV v\dif w.
\end{equation}
This results in the following equation for the density $\tilde\rho$:
\begin{equation}
\label{eq:trho}
\parderiv{\tilde\rho}t+\nabla\cdot(\tilde\rho\tu)=(\rho u_z)
|_{z=0}=0,
\end{equation}
where the latter identity is in effect since the bottom of the domain
is impenetrable and thus $u_z|_{z=0}=0$. Observe that there is no
advective coupling to $u_z$, and only the horizontal momentum is
present in the advection term of \eqref{eq:trho}. Thus, it suffices to
include only the horizontal momentum at the next level of the moment
hierarchy. To obtain the transport equation for the horizontal
momentum, we integrate \eqref{eq:vlasov_g_tf} over $\BV v\dif\BV v\dif
w$:
\begin{equation}
\label{eq:tmom2}
\parderiv{(\tilde\rho\tu)}t+\nabla\cdot(\tilde\rho\tu^2+\tP)=\int\BV
vw f|_{z=0}\dif\BV v\dif w-\nabla\int_0^\infty\bar\phi\dif z,
\end{equation}
where the horizontal pressure tensor $\tP$ is given via
\begin{equation}
\tP=\int(\BV v-\tu)^2\tilde f\dif\BV v\dif w.
\end{equation}
The bottom boundary term in \eqref{eq:tmom2} can be expressed via
\begin{multline}
\int\BV vwf|_{z=0}\dif\BV v\dif w=\int(\BV v-\BV u|_{z=0})w f|_{z=0}
\dif\BV v\dif w+\BV u|_{z=0}\int wf|_{z=0}\dif\BV v\dif w=\\=\int(\BV
v-\BV u|_{z=0}) (w-u_z|_{z=0})f|_{z=0}\dif\BV v\dif w=\BV S_{\BV xz}
|_{z=0},
\end{multline}
where we use the fact that $u_z|_{z=0}=0$. Here, $\BV S_{\BV x z}|_{
  z=0}$ denotes the vector of the surface shear stress:
\begin{equation}
  \BV S_{\BV x z}|_{z=0}=\left.\int\left(\begin{array}{c}(v_x-u_x)
    (w-u_z) \\ (v_y-u_y)(w-u_z) \end{array}\right)f\dif\BV v\dif
  w\right|_{z=0}.
\end{equation}
To obtain the transport equation for the horizontal pressure tensor
$\tP$, we integrate \eqref{eq:vlasov_g_tf} over $\BV v^2\dif\BV v\dif
w$, which leads to the energy equation:
\begin{multline}
\parderiv{}t(\tilde\rho\tu^2+\tP)+\nabla\cdot\big(\tilde\rho\tu^3+\tu
\tP+(\tu\tP)^T+(\tu\tP)^{TT}\big)+\nabla\cdot\int(\BV v-\tu)^3\tilde f
\dif\BV v\dif w=\\=\int\BV v^2w f|_{z=0}\dif\BV v\dif w-\int_0^\infty
\big(\nabla \bar\phi\BV u^T+\BV u\nabla\bar\phi^T\big)\dif z.
\end{multline}
In order to isolate the equation for the horizontal pressure tensor
$\tP$ alone, we observe that, from the horizontal momentum equation
\eqref{eq:tmom2},
\begin{multline}
\parderiv{(\tilde\rho\tu^2)}t=\parderiv{(\tilde\rho\tu)}t\tu^T+\tu
\parderiv{(\tilde\rho\tu^T)}t-\tu^2\parderiv{\tilde\rho}t=\bigg(\int\BV
vw f|_{z=0}\dif\BV v\dif w-\int_0^\infty\nabla\bar\phi\dif z\bigg)\tu^T
+\\+\tu\bigg(\int\BV vw f|_{z=0}\dif\BV v\dif w-\int_0^\infty\nabla\bar
\phi\dif z\bigg)^T+\nabla\cdot(\tilde\rho\tu)\tu^2 -\nabla\cdot(\tilde
\rho\tu^2+\tP)\tu^T-\tu\nabla\cdot(\tilde\rho\tu^2+\tP)^T.
\end{multline}
Substituting the expression for the time derivative and noting that
\begin{equation}
\nabla\cdot(\tilde\rho\tu)\tu^2-\nabla\cdot(\tilde \rho\tu^2)\tu^T
-\tu\nabla\cdot(\tilde \rho\tu^2)^T=-\nabla\cdot(\tilde\rho\tu^3),
\end{equation}
we arrive at the horizontal pressure tensor equation of the form
\begin{multline}
\label{eq:tP}
\parderiv\tP t+\nabla\cdot(\tu \tP)+\tP\nabla\tu+\nabla\tu^T\tP+
\nabla\cdot\int(\BV v-\tu)^3\tilde f \dif\BV v\dif w=\\=\int(\BV v-
\tu)^2 w f|_{z=0}\dif\BV v\dif w-\int_0^\infty\big(\nabla\bar\phi(\BV
u-\tu)^T+(\BV u-\tu)\nabla\bar\phi^T\big)\dif z.
\end{multline}
The bottom boundary term in the above equation can be expressed as
\begin{multline}
\int(\BV v-\tu)^2wf|_{z=0}\dif\BV v\dif w=\int(\BV v-\BV u|_{z=0}+\BV
u|_{z=0}-\tu)^2wf|_{z=0}\dif\BV v\dif w=\\=\int(\BV v-\BV u|_{z=0})^2
wf|_{z=0}\dif\BV v\dif w+\int(\BV v-\BV u|_{z=0})(\BV u|_{z= 0}-\tu)^T
wf|_{z=0}\dif\BV v\dif w+\\+\int(\BV u|_{z=0}-\tu)(\BV v-\BV u|_{z=0})
^Twf|_{z=0}\dif\BV v\dif w +\int(\BV u|_{z=0}-\tu)^2wf|_{z=0}\dif\BV v
\dif w=\\=\BM Q_{\BV x^2z}|_{z=0}+\BV S_{\BV xz}|_{z=0}(\BV u|_{z=0}-
\tu)^T+(\BV u|_{z=0}-\tu)\BV S_{\BV xz}|_{z=0}^T,
\end{multline}
where $\BM Q_{\BV x^2z}|_{z=0}$ is the surface skewness matrix, given
via
\begin{multline}
\BM Q_{\BV x^2z}|_{z=0}=\left.\int(\BV v-\BV u)^2 wf\dif\BV v\dif
w\right|_{z=0}=\\=\left.\int\left(\begin{array}{c|c}(v_x-u_x)^2(w-u_z)
  & (v_x-u_x)(v_y-u_y)(w-u_z) \\\hline (v_x-u_x)(v_y-u_y)(w-u_z) &
  (v_y-u_y)^2(w-u_z)\end{array}\right)f\dif\BV v\dif w\right|_{z=0}.
\end{multline}
At this point, we define the pressure $\tilde p$ in a standard way,
that is, as a normalized trace of $\tP$, and denote the corresponding
horizontal shear stress deviator via $\tS$:
\begin{equation}
\tilde p=\frac 12\trace(\tP),\qquad\tS=\tP-\tilde p\BM I.
\end{equation}
Next, we obtain the equation for $\tilde p$ by taking the normalized
trace of the equation for $\tP$ in \eqref{eq:tP}:
\begin{multline}
\parderiv{\tilde p}t+\nabla\cdot(\tilde p\tu)+\tilde p\nabla\cdot\tu+
\tS:\nabla\tu+\nabla\cdot\tq=\\=q_z|_{z=0}+\BV S_{\BV xz}|_{z=0}\cdot
(\BV u|_{z=0}-\tu)-\int_0^\infty(\BV u-\tu)\cdot\nabla\bar\phi\dif z.
\end{multline}
Above, $\tq$ and $q_z$ are the horizontal and surface heat fluxes,
respectively:
\begin{equation}
\tq=\frac 12\int\|\BV v-\tu\|^2(\BV v-\tu)\tilde f \dif\BV v\dif
w,\quad q_z|_{z=0}=\frac 12\left.\int\|\BV v-\BV u\|^2(w-u_z) f\dif
\BV v\dif w\right|_{z=0}.
\end{equation}
The equation for the horizontal shear stress $\tS$ is obtained by
subtracting the pressure equation from \eqref{eq:tP}:
\begin{multline}
\label{eq:tS}
\parderiv\tS t+\nabla\cdot(\tu\tS)+\tS \nabla\tu+\nabla\tu^T\tS-(\tS:
\nabla\tu)\BM I+\tilde p(\nabla\tu+\nabla\tu^T-(\nabla\cdot\tu)\BM I)+
\\+\nabla\cdot\int(\BV v-\tu)^3\tilde f\dif\BV v\dif w -(\nabla\cdot
\tq)\BM I=\BM Q_{\BV x^2z}|_{z=0}-q_z|_{z=0}\BM I+\\+\BV S_{\BV xz}|_{
  z=0}(\BV u|_{z=0}-\tu)^T+(\BV u|_{z=0}-\tu)\BV S_{\BV xz}|_{z=0}^T-(
\BV S_{\BV xz}|_{z=0}\cdot(\BV u|_{z=0}-\tu))\BM I-\\-\int_0^\infty
\big(\nabla\bar\phi(\BV u-\tu)^T+(\BV u-\tu)\nabla \bar\phi^T-((\BV u-
\tu)\cdot\nabla\bar\phi)\BM I\big)\dif z.
\end{multline}
Following \citep{Abr22}, we split the horizontal skewness tensor into
the appropriate combination of the horizontal heat fluxes and the
traceless deviator $\tQ$:
\begin{subequations}
\begin{equation}
\int(\BV v-\tu)^3\tilde f \dif\BV v\dif w=\frac 12\big(\tq\BM I+(\tq
\BM I)^T+(\tq\BM I)^{TT}\big)+\tQ,
\end{equation}
\begin{equation}
\nabla\cdot\int(\BV v-\tu)^3\tilde f\dif\BV v\dif w-(\nabla\cdot\tq)
\BM I=\frac 12\big(\nabla\tq+\nabla\tq^T-(\nabla\cdot\tq)\BM I\big)+
\nabla\cdot\tQ.
\end{equation}
\end{subequations}
In the same fashion as in our recent work \cite{Abr22}, here we assume
that both horizontal and surface shear stresses are negligible in
comparison with the rest of the terms, which indicates a high Reynolds
number flow. For the horizontal shear stress $\tS$ to remain small,
we, first, impose the balance of external forces and boundary effects
in its equation \eqref{eq:tS}, that is,
\begin{equation}
\BM Q_{\BV x^2z}|_{z=0}=\int_0^\infty \big(\nabla\bar\phi (\BV
u-\tu)^T+(\BV u-\tu)\nabla \bar\phi^T\big)\dif z,
\quad
q_z|_{z=0} =\int_0^\infty (\BV u-\tu)\cdot\nabla\bar\phi\dif z,
\end{equation}
where the latter equation is the trace of the former. Second, we also
impose the balance between the pressure-velocity terms and the heat
fluxes in the equation \eqref{eq:tS} for the horizontal shear stress,
\begin{equation}
\tilde p(\nabla\tu+\nabla\tu^T)+\frac 12(\nabla\tq +\nabla\tq^T)+
\nabla\cdot\tQ=\BM 0,\qquad\tilde p\nabla\cdot\tu+\frac 12\nabla\cdot
\tq=0,
\end{equation}
where, again, the latter equation is the trace of the former.
Substituting the above relations into the pressure equation, we arrive
at
\begin{equation}
\label{eq:tp}
\parderiv{\tilde p}t+\tu\cdot\nabla\tilde p=0,
\end{equation}
that is, in the same manner as in \citep{Abr22}, the pressure is
preserved along the stream lines (balanced flow). For the vanishing
shear stress, the momentum equation in \eqref{eq:tmom2} becomes
\begin{equation}
\label{eq:tmom}
\parderiv{(\tilde\rho\tu)}t+\nabla\cdot(\tilde\rho\tu^2)+\nabla\tilde
p=-\nabla\int_0^\infty\bar\phi\dif z.
\end{equation}
The density equation in \eqref{eq:trho}, together with the pressure
equation \eqref{eq:tp} and the momentum equation \eqref{eq:tmom},
constitute the two-dimensional balanced flow equations in the presence
of the gravitational acceleration.

For a numerical computation, it is more convenient to reformulate the
pressure equation \eqref{eq:tp} in the form of a conservation law. For
that, we can use the inverse kinetic temperature as a variable;
indeed, denoting $\tilde\theta=\tilde p/\tilde\rho$, one can, with the
help of \eqref{eq:trho}, rearrange \eqref{eq:tp} in the form
\begin{equation}
\label{eq:ttheta}
\parderiv{(\tilde\theta^{-1})}t+\nabla\cdot(\tilde\theta^{-1}\tu)=0.
\end{equation}
Together, the equations \eqref{eq:trho}, \eqref{eq:tmom} and
\eqref{eq:ttheta} form the system of nonlinear conservation laws for
the density $\tilde\rho$, momentum $\tilde\rho\tu$ and inverse kinetic
temperature $\tilde\theta^{-1}$, where the latter can be related to
the standard temperature (in $^\circ$K) via an appropriate gas
constant.

\subsection{Approximation for the vertically integrated mean field potential}

To estimate the effect of the potential forcing in the momentum
equation \eqref{eq:tmom}, we assume that the gas particles interact
via the hard sphere mean field potential -- that is $\phi(r)$ is zero
when $r$ is greater than the diameter of the {\em protection sphere}
\citep{Cer2} (that is, the sphere of minimal distance between the
centers of the colliding hard spheres), and becomes infinite for $r$
less than that. In such a case, the spatial integral in
\eqref{eq:bphi} equals the volume of this protection sphere regardless
of the value of the kinetic temperature:
\begin{equation}
\int\left[1-\exp\left(-\frac{\phi(\|\BV y\|)}\theta\right)\right]
\dif\BV y=V_{prot}=8 V_{HS}.
\end{equation}
Above, $V_{prot}$ and $V_{HS}$ denote the volumes of the protection
and hard spheres, respectively, with the latter being eight times
smaller than the former, because the size of the protection sphere is
twice that of any of the two colliding hard spheres. Substituting
$V_{HS}$ into \eqref{eq:bphi} and denoting the density of the hard
sphere via $\rho_{HS}=m/V_{HS}$, we arrive at a rather simple
expression for the hard sphere mean field potential $\bar\phi_{HS}$:
\begin{equation}
\label{eq:bphi_HS}
\bar\phi_{HS}=\frac{4\rho p}{\rho_{HS}}.
\end{equation}
Note that the hard sphere mean field potential in \eqref{eq:bphi_HS}
introduces the same correction into the momentum equation in
\eqref{eq:bflow3D} as does the Enskog correction in the Enskog--Euler
equations (see \citet{Abr17} for details).

Next, we assume that, in the vertical direction, the hydrostatic
balance is achieved:
\begin{equation}
\parderiv pz=-g\rho=-\frac{gp}\theta,\qquad\parderiv{\ln p}z=-\frac g
\theta,\qquad p=p|_{z=0}\exp\bigg(-g\int_0^z\frac{\dif\xi}{\theta(\xi)
}\bigg).
\end{equation}
We now express $\rho$ from the hydrostatic balance relation and write
\begin{equation}
\rho p=-\frac p g\parderiv pz=-\frac 1{2g}\parderiv{(p^2)}z=\frac{
  (p|_{z=0})^2}\theta\exp\bigg(-2g\int_0^z\frac{\dif\xi}{\theta(\xi)}
\bigg).
\end{equation}
For the integral of $\rho p$ in the vertical direction we thus have
\begin{equation}
\int_0^\infty\rho p\dif z=(p|_{z=0})^2\int_0^\infty\exp\bigg(-2g
\int_0^z\frac{\dif\xi}{\theta (\xi)}\bigg)\frac{\dif z}{\theta(z)}
=\frac{(p|_{z=0})^2}{2g}\int_0^\infty e^{-\eta}\dif\eta=\frac{
  (p|_{z=0})^2}{2g}.
\end{equation}
Above, the variable $\eta=\eta(z)$ is given via
\begin{equation}
\eta(z)=2g\int_0^z\frac{\dif\xi}{\theta(\xi)},
\end{equation}
and we assume that $\theta$ remains bounded above and below with
altitude (which is indeed the case in the troposphere and
stratosphere). For the vertical integral of the hard sphere mean field
potential $\bar\phi_{HS}$ we thus have
\begin{equation}
\int_0^\infty\bar\phi_{HS}\dif z=\frac 4{\rho_{HS}}\int_0^\infty\rho
p\dif z=\frac{2(p|_{z=0})^2}{g\rho_{HS}}.
\end{equation}
What remains to be done is to evaluate the surface pressure
$p|_{z=0}$.  For that, we recall that the surface pressure equals the
total mass of the air column above the surface per unit area,
multiplied by the gravity acceleration, which yields
\begin{equation}
p|_{z=0}=g\int_0^\infty\rho\dif z=g\int\bigg(\int_0^\infty f\dif
z\bigg)\dif\BV v\dif w=g\int\tilde f\dif\BV v\dif w=g\tilde\rho,
\end{equation}
and, subsequently,
\begin{equation}
\int_0^\infty\bar\phi_{HS}\dif z=\frac{2g\tilde\rho^2}{\rho_{HS}}.
\end{equation}
As a result, the horizontal momentum equation for the hard sphere gas
becomes
\begin{equation}
\label{eq:tmom_HS}
\parderiv{(\tilde\rho\tu)}t+\nabla\cdot(\tilde\rho\tu^2)+\nabla
\left(\tilde p+\frac{2g\tilde\rho^2}{\rho_{sp}}\right)=\BV 0.
\end{equation}

\section{Numerical simulations}
\label{sec:numerical}

Here we present the results of numerical simulations of a balanced
two-dimensional large scale gas flow for two scenarios -- an inertial
jet, and a cyclostrophic vortex. In both scenarios, we recreate
(rather crudely) the main parameters of Earth's atmosphere.  In
particular, the acceleration constant in \eqref{eq:tmom_HS} is set to
$g=9.81$ m/s$^2$, which corresponds to Earth's gravity, while the
vertically integrated density $\tilde\rho$ and temperature
$\tilde\theta$ of the gas are chosen to correspond to those typically
observed in large scale atmospheric flows.

\subsection{Density of the air-like hard sphere gas}

The Earth's atmosphere consists of air, which is a mixture of various
gases -- primarily diatomic, such as the molecular nitrogen N$_2$ and
molecular oxygen O$_2$, but with a small amount of polyatomic gases,
such as water H$_2$O and carbon dioxide CO$_2$. Our theory is,
however, developed for a monatomic hard sphere gas. Thus, in order to
numerically simulate the behavior of the Earth atmosphere using our
model, we need to ``create'' a hypothetical hard sphere gas with a
density which matches known physical properties of air. Here, we
choose the two properties of our hard sphere gas to match with the air
-- the molar mass, and the viscosity.

As follows from our gas model above, the equations in \eqref{eq:trho},
\eqref{eq:tp}, \eqref{eq:ttheta} and \eqref{eq:tmom_HS} are derived
from the assumption that the molecules interact in a fully
time-reversible fashion. On the contrary, in the standard kinetic
theory, the viscosity properties of a hard sphere gas are obtained
directly from the time-irreversible Boltzmann collision integral
\citep{ChaCow,Gra,Cer,HirCurBir}. This is not necessarily a
contradiction -- clearly, in a real gas, both types of interactions
are present (for example, while potential interactions are
time-reversible, the quantum-mechanical Pauli repulsion is inherently
stochastic), and can apparently manifest at different spatial
scales. In particular, turbulence is not observed in flows with high
Knudsen number (such as thin channels and capillaries) where the gas
motion is dominated by viscous effects, whereas the situation reverses
itself at low Knudsen numbers. Thus, the same gas can be treated in
the context of the time-reversible model at large scales, and as a
time-irreversible process at viscous scales.

Treating the collisions as time-irreversible at viscous scales, we
recall the well known formula for the dynamic viscosity of the hard
sphere gas \citep{Gra,ChaCow,HirCurBir}:
\begin{equation}
\mu=\frac 5{16\sqrt\pi}\frac m{\sigma^2}\sqrt\theta=\frac
5{16\sqrt\pi}\frac M{N_A\sigma^2}\sqrt\theta,
\end{equation}
where $\sigma$ is the diameter of the hard sphere, $N_A=6.02214\cdot
10^{23}$ mol$^{-1}$ is the Avogadro number, and $M$ is the molar mass
of the gas. In order to relate the expression above to the density of
the hard sphere $\rho_{HS}$, we observe that
\begin{equation}
\theta=\frac{RT}M,\qquad\rho_{HS}=\frac m{V_{HS}}=\frac{6m}{\pi
  \sigma^3} =\frac{6M}{\pi N_A \sigma^3},
\end{equation}
where $R=8.31446$ kg m$^2$/K mol s$^2$ is the universal gas constant,
and $T$ is the usual temperature in $^\circ$K. Expressing $\sigma$ via
$\rho_{HS}$ from the latter equation, and substituting the expression
for $\theta$, we obtain the formula for $\rho_{HS}$ via the viscosity
$\mu$ and temperature $T$:
\begin{equation}
\label{eq:rho_HS_mu_T}
\rho_{HS}=\frac{384}{5\sqrt 5}\bigg(\frac{N_A^2M\mu^6}{\pi R^3T^3}
\bigg)^{1/4}.
\end{equation}
With \eqref{eq:rho_HS_mu_T}, we can ``create'' a theoretical hard
sphere analog of any gas with a prescribed molar mass $M$ and known
viscosity $\mu$ at a given temperature $T$. For air, $M=2.897\cdot
10^{-2}$ kg/mol, and we take $\mu=1.8194\cdot 10^{-5}$ kg/m s at
$T=293.15$ K \citep{Jon}, for which \eqref{eq:rho_HS_mu_T} yields
$\rho_{HS}=1850$ kg/m$^3$. This value is used throughout all
computations below.

\subsection{Two simulated scenarios of a balanced flow}

Below we study the following two special cases of a balanced flow:
\begin{itemize}
\item {\bf Inertial jet.} In an inertial jet, the pressure is constant
  throughout the domain. Observe that, in our equations, the Coriolis
  acceleration of Earth rotation is not present, which corresponds
  roughly to the equatorial region. Thus, such inertial flow describes
  a special case of {\em geostrophic flow} near the equator.
\item {\bf Cyclostrophic vortex.} In a cyclostrophic vortex, the
  centripetal force, acting on a rapidly rotating gas, is balanced by
  the pressure gradient, and the velocity is orthogonal to both.
  Again, given the absence of the Coriolis acceleration, this scenario
  corresponds to a large scale rapidly rotating flow, where the
  Coriolis acceleration is negligible in comparison with the
  centripetal acceleration -- that is, a fully developed tropical
  cyclone.
\end{itemize}

\subsection{Computation of initial and boundary conditions}
\label{sec:IC_BC}

Below we describe how the initial and boundary values of the velocity,
temperature and pressure are specified in each simulated scenario.

\begin{itemize}
\item First, we specify the velocity $\tu$ of the flow in the domain,
  and on those boundaries, where the Dirichlet boundary condition is
  specified. The way we specify the velocity is entirely
  scenario-dependent; in the inertial jet scenario, the initial
  velocity field forms a jet stream, while in the cyclostrophic vortex
  scenario the initial velocity field forms a rotating flow with an
  appropriate radial profile.
\item Once $\tu$ is defined, we compute the kinetic temperature
  $\tilde\theta$ of the flow using the Bernoulli law for a
  compressible gas:
  \begin{equation}
  \tilde\theta=\frac{R\tilde T_0}M-\frac{\gamma-1}{2\gamma}\|\tu\|^2,
  \end{equation}
  where the background temperature of the gas is set to $\tilde
  T_0=250$ K, that is, the approximate vertically averaged temperature
  of air throughout the troposphere.  Observe, however, that in
  present setting the gas is two-dimensional, and, therefore, the
  effective adiabatic exponent $\gamma=2$.
\item Once $\tu$ and $\tilde\theta$ are defined, we specify the
  pressure $\tilde p$ (which automatically yields the density via
  $\tilde\rho=\tilde p/\tilde\theta$). Below, we study two scenarios:
  an inertial jet, and a cyclostrophic vortex. In the inertial jet
  scenario, we set the pressure to a constant value $\tilde p_0$,
   \begin{equation}
    \tilde p_0=\frac{R\tilde\rho_0\tilde T_0}M,
  \end{equation}
  where $\tilde\rho_0=10^4$ kg/m$^2$, which constitutes the
  approximate mass of the air column per unit area of the Earth
  surface. In the cyclostrophic vortex scenario, the pressure gradient
  is balanced by the centripetal force:
  \begin{equation}
    \nabla\tilde p=-\tilde\rho\BV\Omega\times(\BV\Omega\times\BV r)=
    \tilde\rho\|\BV\Omega\|^2\BV r,
  \end{equation}
  where $\BV r$ is the coordinate of the point relative to the center
  of rotation, $\BV\Omega$ is the angular velocity of rotation, and
  the latter identity is valid for a planar rotation (that is, when
  $\BV r\perp\BV\Omega$). Here, the velocity $\tu$ is orthogonal to
  both $\BV r$ and $\BV\Omega$, that is,
  \begin{equation}
  \tu=\BV\Omega\times\BV r,\qquad\text{and, since }\BV
  r\perp\BV\Omega,\qquad \|\tu\|=\|\BV\Omega\|\|\BV r\|.
  \end{equation}
  Upon substitution, this leads to
  \begin{equation}
  \nabla\tilde p=\frac{\tilde\rho\|\tu\|^2}{\|\BV r\|^2}\BV r.
  \end{equation}
  Assuming, in turn, that $\tilde\rho$, $\tu$ and $\tilde p$ do not
  depend on the angle, and only depend on $\|\BV r\|$, we denote
  $r=\|\BV r\|$, $\tilde u= \|\tu\|$, and arrive at the following
  explicit formula for $\tilde p$:
  \begin{equation}
  \label{eq:p_cyclostrophic}
  \deriv{\tilde p}r=\frac{\tilde\rho\tilde u^2}r=\frac{\tilde p\tilde
    u^2}{\tilde\theta r},\qquad\text{or}\qquad\tilde p(r)=\tilde p_0
  \exp\left(-\int_r^\infty\frac{\tilde u^2(\xi)}{\tilde\theta(\xi)}
  \frac{\dif\xi}\xi\right).
  \end{equation}
  Above, in the integration formula we assume that $\tilde u$ vanishes
  sufficiently far away from the center of rotation. Also, in
  practice, one can simplify the integration by presuming that
  $\tilde\theta$ varies weakly in comparison to $\tilde u^2$ under the
  integral, and factor the kinetic temperature out of the integration
  (as we do further below).
\end{itemize}

\subsection{Numerical methods and software}

In the current work, we use the same software as in our previous work
\citep{Abr22} -- namely, we use OpenFOAM \citep{WelTabJasFur} to
perform all numerical simulations. Noting that the equations
\eqref{eq:trho}, \eqref{eq:ttheta} and \eqref{eq:tmom_HS}, for the
density, inverse temperature and momentum, respectively, comprise a
system of nonlinear conservation laws, we simulate them with the help
of an appropriately modified \texttt{rhoCentralFoam} solver
\citep{GreWelGasRee}, which uses the central scheme of \citet{KurTad}
for the numerical finite volume discretization, with the flux limiter
due to \citet{vanLee}. The time-stepping of the method is adaptive,
based on the 20\% of the maximal Courant number.

\subsection{Numerical simulation of an inertial jet}

In the inertial jet scenario, we simulate a large scale jet stream in
a channel-like domain. The spatial domain is 2500 km in length, and
500 km in width. The spatial discretization is uniform in both
directions, with a step of 2.5 km, which constitutes 1000 cells in the
zonal direction, and 200 cells in the meridional direction. The domain
has a 100 km-wide inlet in the middle of the western wall, while the
outlet is the whole 500 km-wide eastern boundary. The initial velocity
field is given via the shear jet
\begin{equation}
\label{eq:u_inertial}
\tu|_{t=0}=(u(y/d),0),\qquad u(z)=u_0\frac{1+\cos(\pi z)}2(1-\alpha
z),
\end{equation}
where the vertical coordinate $y$ is given relative to the central
axis of the channel, $d=50$ km, and $u_0=35$ m/s. As we can see, the
speed of the flow is 35 m/s in the middle of the channel, and smoothly
decays to zero at the distance of 50 km both to the north and south
(which also corresponds to the boundaries of the inlet). The weak
asymmetry parameter $\alpha=0.05$ is introduced in the same manner as
by \citet{McCHai}, to break up possible mirror symmetry effects of the
flow in the two-dimensional domain. Observe that the typical kinematic
viscosity of air at normal conditions is $\sim 10^{-5}$ m$^2$/s, the
characteristic size of our flow is no less than $10^5$ m (the width of
the jet), and the reference velocity is no less than $10$ m/s, which
yields the value of the Reynolds number $\Rey\sim 10^{11}$. Therefore,
the assumptions made in the course of the derivation of our model are
clearly valid for this set-up.

The boundary conditions are specified as follows. At the inlet, the
velocity is set to \eqref{eq:u_inertial} at all times, with the rest
of the variables following the procedure set forth in Section
\ref{sec:IC_BC}. At the outlet, all variables are set to the free
outflow conditions (that is, zero normal gradient). At the walls,
which comprise the remainder of the domain boundary, the velocity is
set to the no-slip condition, while the rest of the variables is set
to the zero normal gradient condition (which corresponds to zero
momentum flux for the pressure, and zero heat flux for the
temperature).

We emphasize that these initial and boundary conditions correspond to
a steady state for the conventional Euler equations, and even for our
balanced flow equations taken without the mean field forcing in the
momentum equation \eqref{eq:tmom} (or its special case
\eqref{eq:tmom_HS} for a hard sphere gas). Thus, any observed
non-steady effects originate from the presence of the mean field
forcing due to the molecular interaction potential.

\begin{figure}
\includegraphics[width=\textwidth]{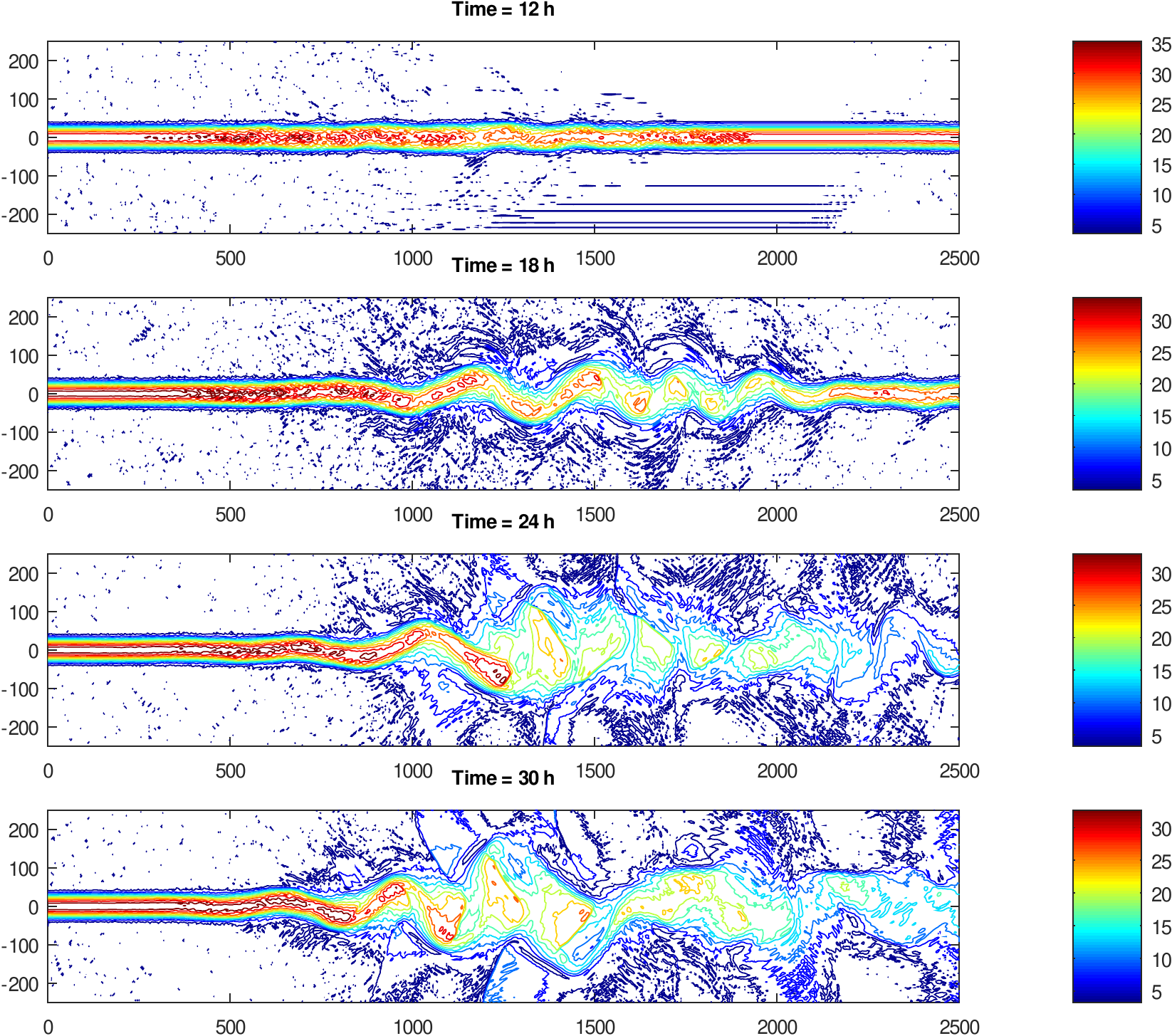}
\caption{Speed of the inertial flow, captured at 12, 18, 24 and 36
  hours of elapsed model time.}
\label{fig:u_inertial}
\end{figure}

Starting from the initial conditions described above, we integrate the
system of equations in \eqref{eq:trho}, \eqref{eq:ttheta} and
\eqref{eq:tmom_HS} forward for 72 hours in model time. The time step,
chosen from the CFL criterion, varied between 12 and 16 seconds. In
Figure~\ref{fig:u_inertial}, we show the speed $\|\tu\|$ of the flow,
captured at 12, 18, 24 and 30 hours elapsed model time, on contour
plots. Here we can see that small fluctuations in the jet manifest at
12 hours. At 18 hours, the jet visibly meanders roughly between 1000
and 2000 km. At 24 and 30 hours, the jet is completely broken up in
the second half of the domain. In fact, visually, the snapshots of the
flow speed at 24 and 30 hours resemble the famous Reynolds experiment
\citep{Rey}. After 30 hours, the general configuration of the flow
does not seem to visibly change any further, and thus we do not show
the snapshots past that time.

\begin{figure}
\includegraphics[width=\textwidth]{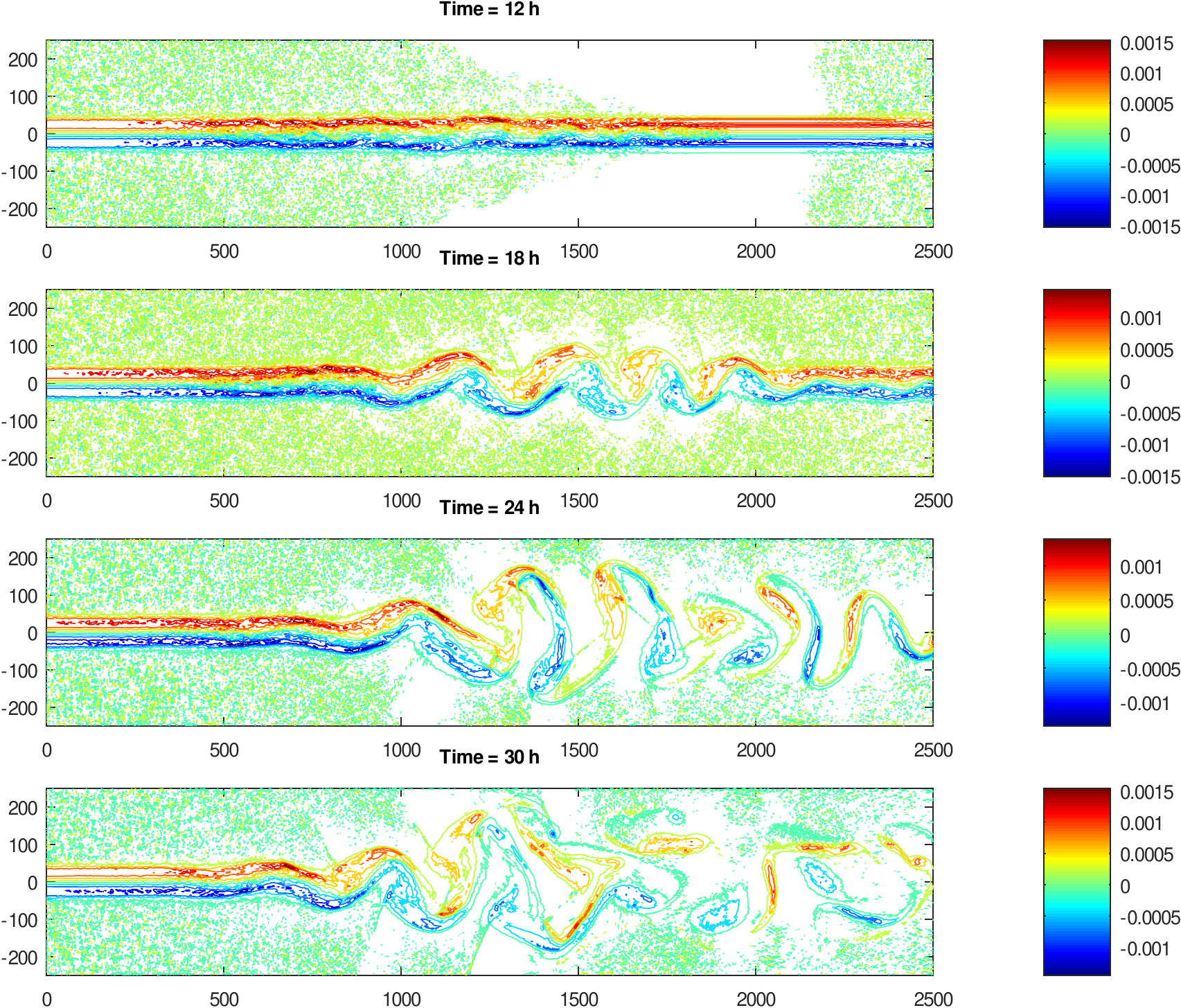}
\caption{Vorticity of the inertial flow, captured at 12, 18, 24 and 36
  hours of elapsed model time.}
\label{fig:vort_inertial}
\end{figure}

The vorticity $\tomega$ is given via the curl of velocity $\tu$:
\begin{equation}
\tomega=\nabla\times\tu.
\end{equation}
Since $\tu$ is confined to the $xy$-plane, $\tomega$ is orthogonal to
this plane, and only has the $z$-component, given via
\begin{equation}
\tilde\omega_z=\parderiv{\tilde u_y}x-\parderiv{\tilde u_z}y.
\end{equation}
In Figure~\ref{fig:vort_inertial}, we show the contour plots of
$\tilde\omega_z$ for the inertial flow, also captured at 12, 18, 24
and 30 hours elapsed model time. The situation here is similar to what
was observed in Figure~\ref{fig:u_inertial} for the speed of the
inertial flow; namely, small vorticity fluctuations are observed at 12
hours, and visible meandering manifests at 18 hours. At 24 hours, the
structure of vorticity resembles a K\'arm\'an vortex trail
\citep{vKar}; at 30 hours the structure of the vorticity is completely
lost in the second half of the channel, and the motion appears to be
fully chaotic.

\begin{figure}
\includegraphics[width=\textwidth]{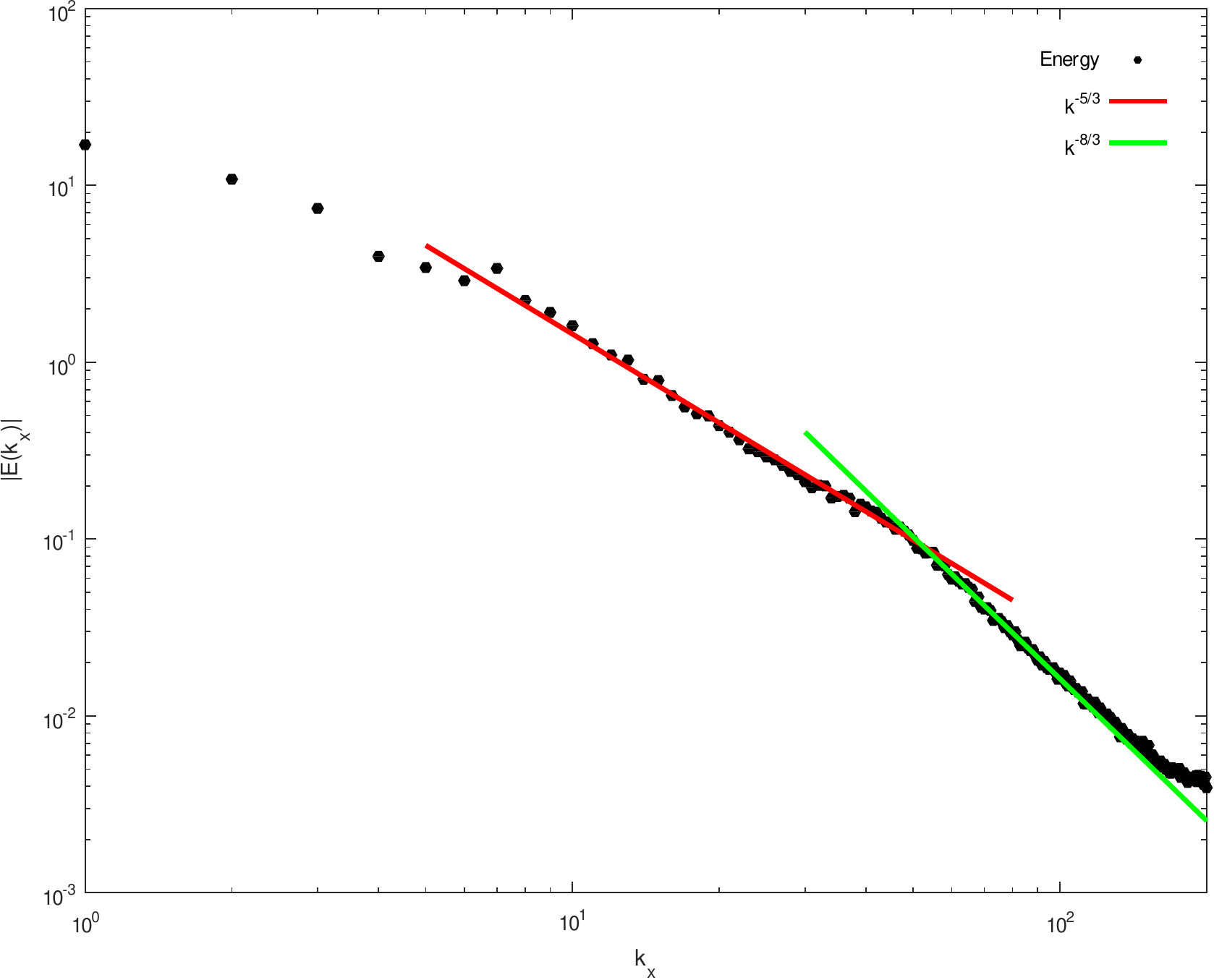}
\caption{Time-averaged kinetic energy spectrum of the inertial flow.
  The straight lines, corresponding to $k^{-5/3}$ and $k^{-8/3}$
  slopes, are provided for reference.}
\label{fig:energy_inertial}
\end{figure}

In Figure~\ref{fig:energy_inertial}, we show the time-averaged kinetic
energy spectrum of the flow, computed in the rectangular sub-region of
the domain, which extends from 1500 to 2500 km zonally, and from
$-100$ to $+100$ km meridionally. This spectrum was computed as in
\citep{Abr22}: first, the kinetic energy $E(\BV x)=\|\BV u(\BV
x)\|^2/2$ was averaged across the channel (thus becoming the function
of the $x$-coordinate only). Then, the linear trend was subtracted
from the result in the same manner as was done by \citet{NasGag}, to
ensure that there was no sharp discontinuity between the energy values
at the western and eastern boundaries of the region. Finally, the
one-dimensional discrete Fourier transformation was applied to the
result. The subsequent time-averaging of the modulus of the Fourier
transform was computed between $t=24$ and $t=72$ hours. As we can see
in Figure~\ref{fig:energy_inertial}, Kolmogorov's famous
$k^{-5/3}$-decay of the kinetic energy spectrum occurs roughly between
the wavenumbers 8 and 50; for smaller scales, the $k^{-8/3}$-decay is
observed.

\begin{figure}
\includegraphics[width=\textwidth]{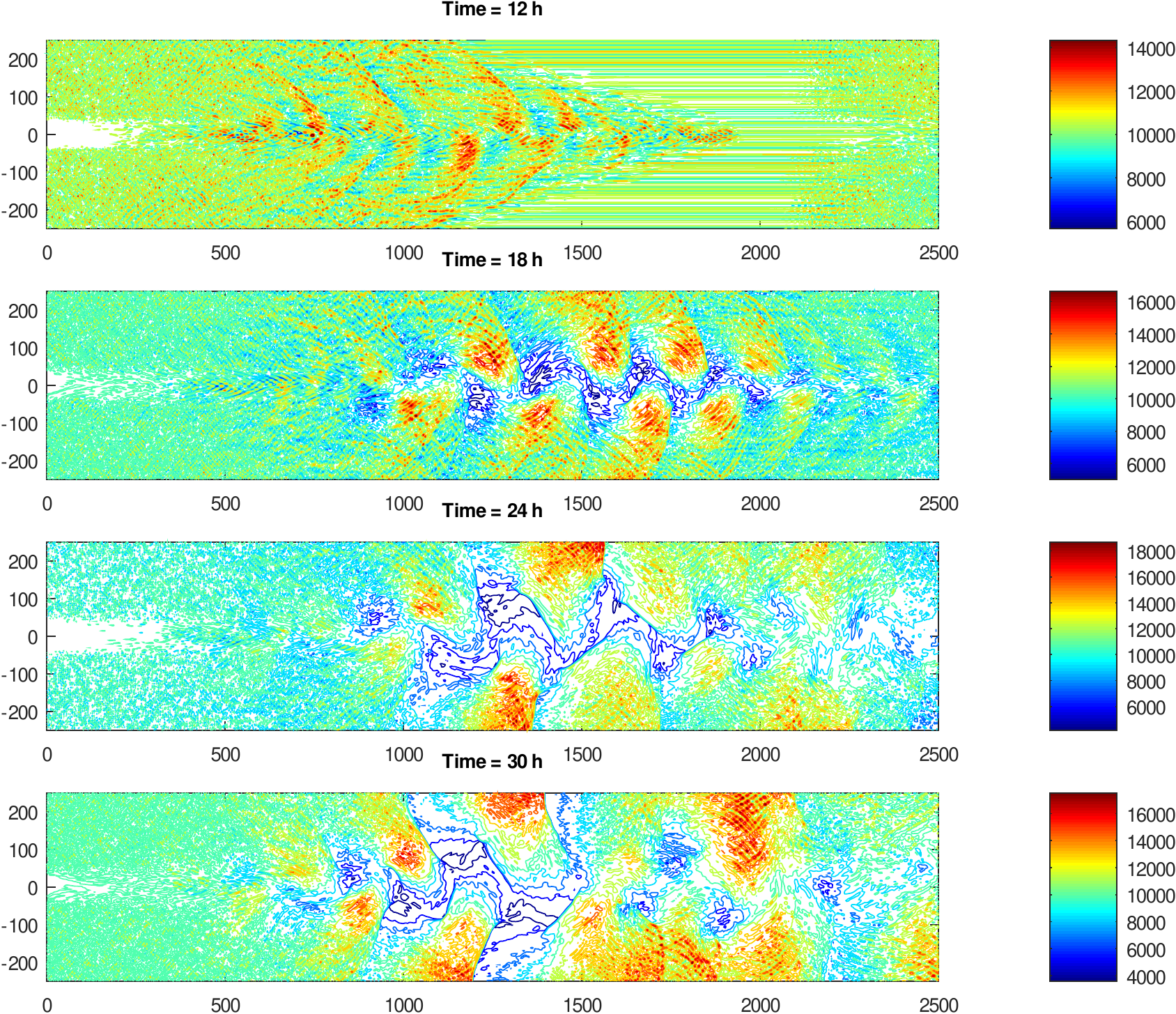}
\caption{Density of the inertial flow, captured at 12, 18, 24 and 36
  hours of elapsed model time.}
\label{fig:rho_inertial}
\end{figure}

Despite promising results in simulating turbulent motions of the
velocity, our model also has its limitations. In particular, for a
fully developed turbulent flow, we found that the density $\tilde\rho$
may attain unrealistic values. In Figure~\ref{fig:rho_inertial}, we
show the contour plots of the density for the same times as the
velocity and vorticity above, namely, at 12, 18, 24 and 30 hours.
Observe that, in a fully developed chaotic flow, the density varies
between $4\cdot 10^3$ and $1.6\cdot 10^4$ kg/m$^2$ (with the
background value $10^4$ kg/m$^2$), which, of course, does not happen
in Earth's atmosphere.

The likely reason for this behavior is that our model is derived under
the assumption that the flow remains balanced unconditionally, while
in reality balanced flows manifest largely in the presence of
relatively small density gradients. Should a large density gradient
develop locally, the flow is likely to become isentropic in that
region, and the usual compressible Euler equations would apply
instead. Thus, in order to correct such a behavior of our model, one
likely needs to introduce an appropriate mechanism of controlled
switching to the compressible Euler equations locally, should the
density gradient become ``too large''. In the present work, however,
we report the results of our simulation without any corrections.

\subsection{Numerical simulation of a cyclostrophic vortex}

\begin{figure}
\includegraphics[width=\textwidth]{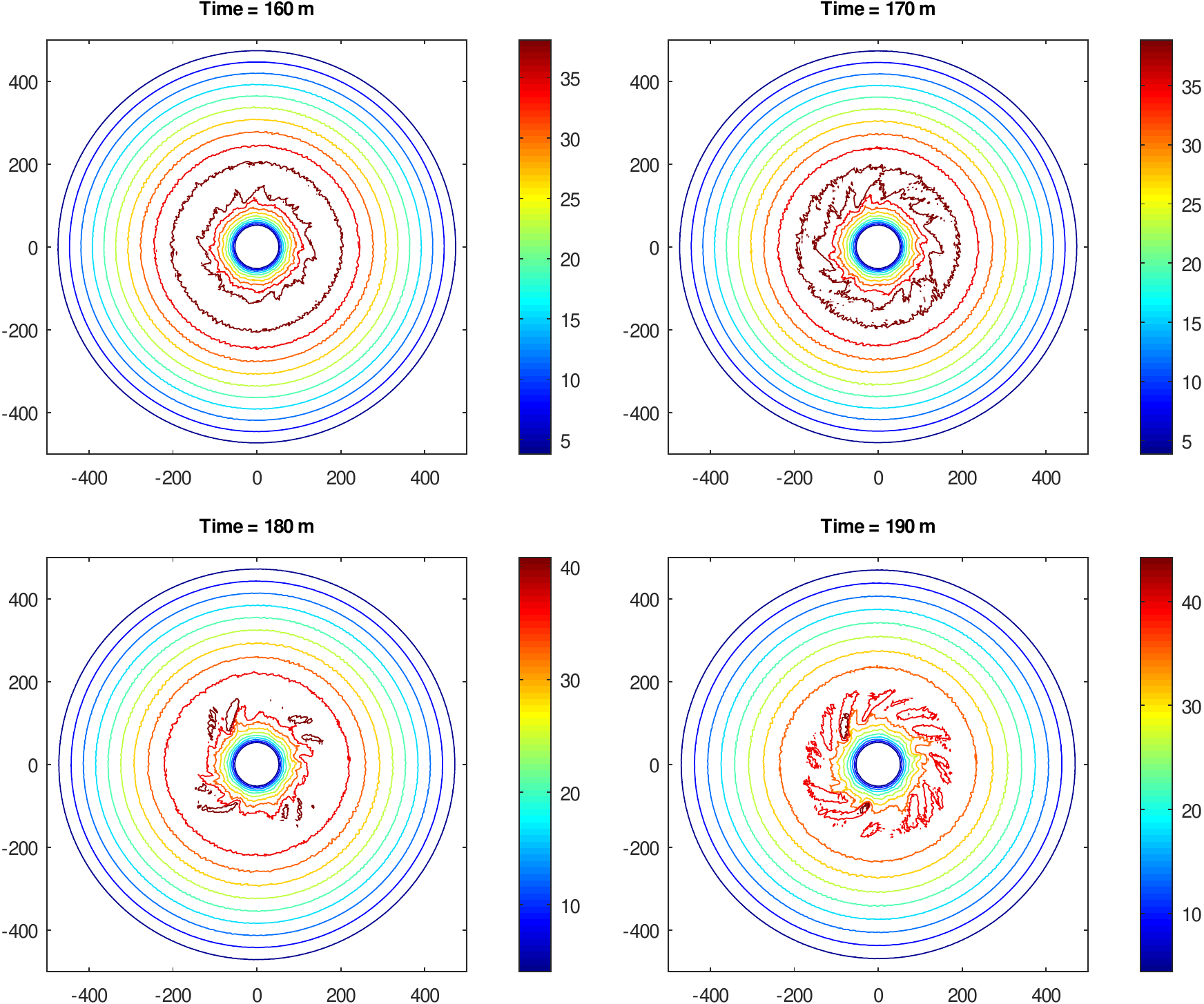}
\caption{Speed of the cyclostrophic flow, captured at 160, 170, 180 and 190
  minutes of elapsed model time.}
\label{fig:u_cyclostrophic}
\end{figure}

\begin{figure}
\includegraphics[width=\textwidth]{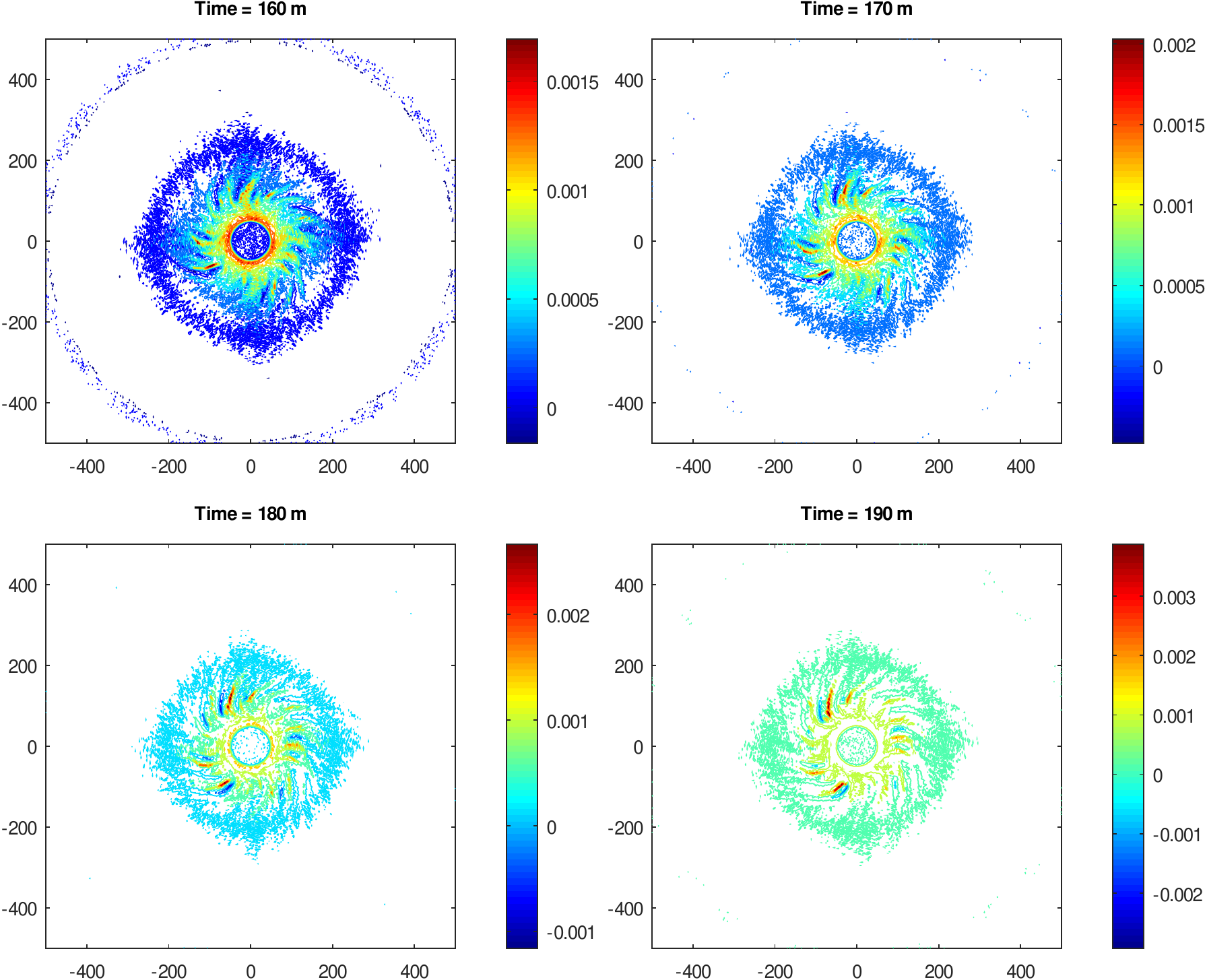}
\caption{Vorticity of the cyclostrophic flow, captured at 160, 170, 180 and 190
  minutes of elapsed model time.}
\label{fig:vort_cyclostrophic}
\end{figure}

In the scenario with a cyclostrophic vortex, we simulate a large scale
rapidly rotating flow, resembling a tropical cyclone, in a square
domain of size 1000$\times$1000 km. The spatial discretization step is
set to 2 km, which constitutes 500 cells in both horizontal
directions. The boundary of the domain is impenetrable with
appropriate boundary conditions, and the counter-clockwise vortex is
confined entirely within it. Naturally, the direction of the velocity
at each point is orthogonal to the direction towards the center of the
vortex, and the speed of the flow is a function of the distance to the
center, but not the angle. As a function of the distance to the center
$r$, we set the speed of the flow to
\begin{equation}
\tilde u(r)=\left\{\begin{array}{l@{\qquad}l}\displaystyle u_0\frac{4
  \ln(r/r_0)\ln(R_0/r)}{\ln^2(R_0/r_0)}, & r_0\leq r\leq R_0,
\\ \text{zero}, & \text{otherwise}.\end{array}\right.
\end{equation}
Here, the maximum speed $u_0=40$ m/s is achieved at the geometric mean
distance $r_{\text{max }u}=\sqrt{r_0R_0}$, where $r_0$ and $R_0$ are
the inner and outer radii of the vortex, respectively. For the
simulation, we set $R_0=10 r_0=500$ km (and thus, $r_{\text{max }
  u}\approx 158$ km). The reason why $\tilde u(r)$ is chosen in this
manner, is that it is, effectively, a function of $\ln r$, which
synergizes well with the computation of the integral in the
cyclostrophic pressure profile formula \eqref{eq:p_cyclostrophic}.
The Reynolds number of this scenario is similar to that of the
inertial flow examined above.

\begin{figure}
\includegraphics[width=\textwidth]{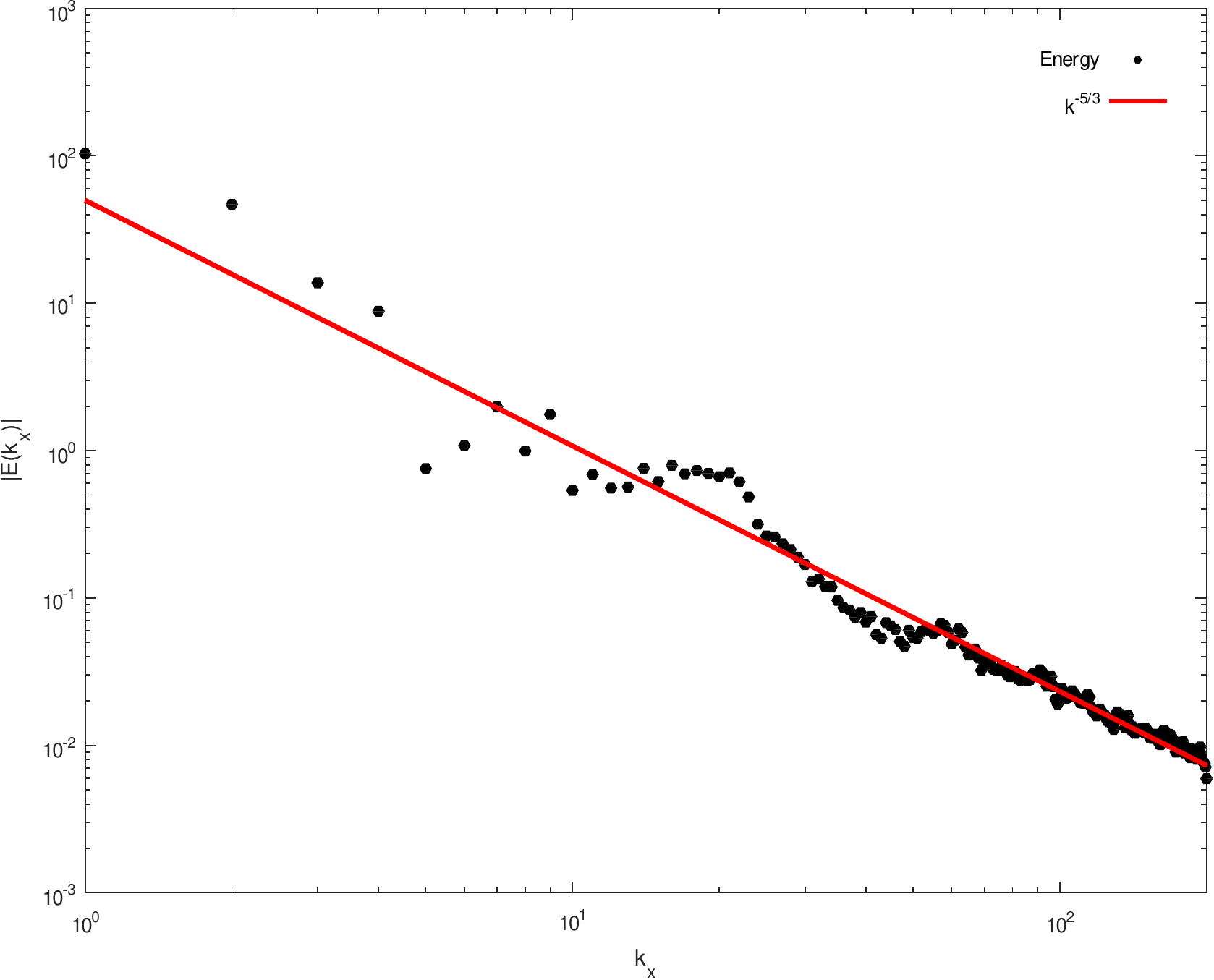}
\caption{The kinetic energy spectrum of the cyclostrophic flow.  The
  straight line, corresponding to $k^{-5/3}$ slope, is provided for
  reference.}
\label{fig:energy_cyclostrophic}
\end{figure}

Upon integrating this initial set-up forward in time, we observed the
following behavior. The time step, chosen from the CFL criterion,
varied between 5 and 6 seconds. The flow remained largely laminar for
the initial 150 minutes of the elapsed model time, after which
turbulent motions rapidly developed around the region of the maximal
flow speed.  Shortly after 200 minutes of the elapsed model time, a
numerical instability occurred in the solution, and a floating point
exception was generated by the software. Thus, turbulent flow was
observed between 150 and 200 minutes of the elapsed time.

A possible reason for the manifestation of the numerical instability
is the apparent lack of any damping effects in the model, similarly to
what we observed in our recent work \citep{Abr21}. In the inertial jet
scenario above, the developing regions with large density gradients
left the domain through the outlet before causing further problems (in
a way, damping was created by the boundary conditions); here,
however, the rotating flow is contained entirely within the domain,
thus allowing instabilities to develop without restrictions.

In Figure~\ref{fig:u_cyclostrophic}, we show the speed $\|\tu\|$ of
the flow, captured at 160, 170, 180 and 190 minutes of the elapsed
model time, on contour plots. Here we can see that small speed
fluctuations manifest at 160 minutes, and become progressively larger
at 170, 180 and 190 minutes. These fluctuations are, however, mainly
confined to the region of maximal velocity of the vortex, with the
flow at the ``outskirts'' remaining laminar.

\begin{figure}
\includegraphics[width=\textwidth]{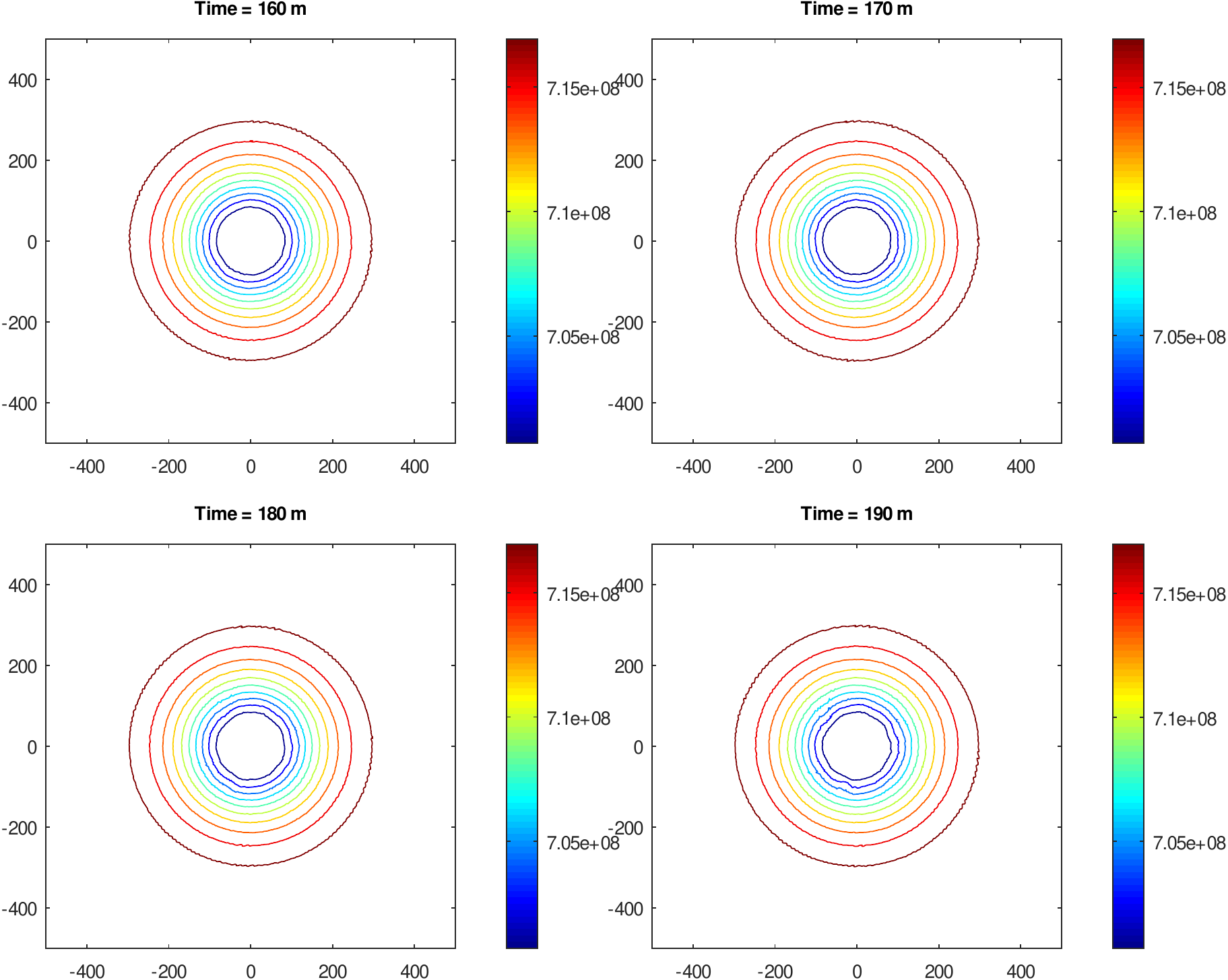}
\caption{Pressure of the cyclostrophic flow, captured at 160, 170, 180 and 190
  minutes of elapsed model time.}
\label{fig:p_cyclostrophic}
\end{figure}

In Figure~\ref{fig:vort_cyclostrophic}, we show the vertical vorticity
component $\tilde\omega_z$ of the flow, captured at 160, 170, 180 and
190 minutes of the elapsed model time, on contour plots. Likewise, one
can observe that the vorticity is largely confined to the region with
maximal flow speed, and its fluctuations become progressively larger
with elapsed time.

\begin{figure}
\includegraphics[width=\textwidth]{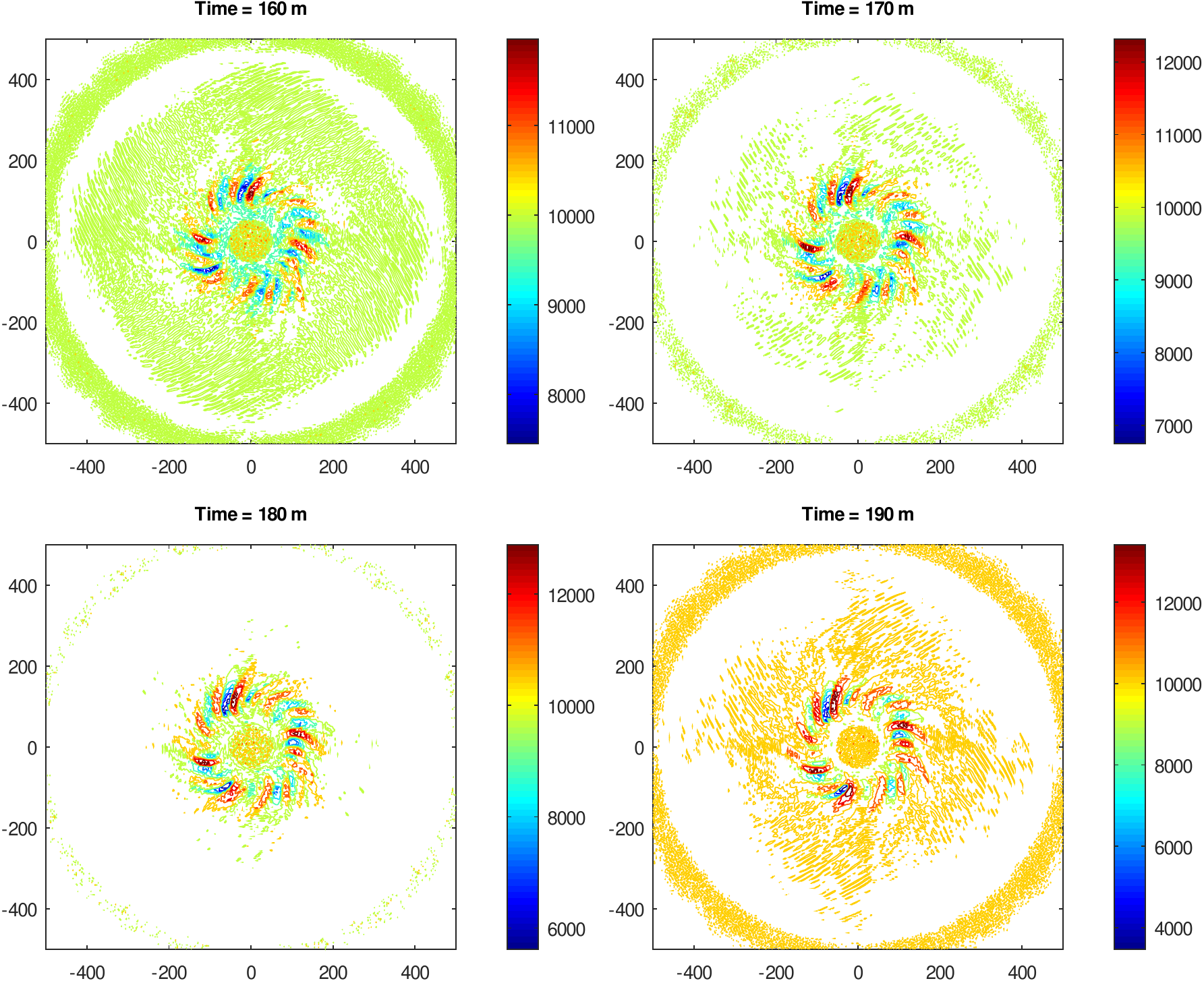}
\caption{Density of the cyclostrophic flow, captured at 160, 170, 180 and 190
  minutes of elapsed model time.}
\label{fig:rho_cyclostrophic}
\end{figure}

In Figure~\ref{fig:energy_cyclostrophic}, we show the time-averaged
kinetic energy spectrum of the flow. This spectrum was computed in the
square sub-region of the domain, which extends from $-400$ to $+400$
km both meridionally and zonally, to reduce the effects of relatively
``calm'' corners of the domain. This computation was done in the same
manner as above for the inertial flow, except, due to isotropy of the
flow, here we computed the spectrum both in the zonal and meridional
directions, and then took the average. The time averaging was done
over the interval between 180 and 200 minutes of the elapsed model
time, during which the most turbulent flow was observed.  As we can
see in Figure~\ref{fig:energy_cyclostrophic}, the kinetic energy
spectrum largely matches Kolmogorov's $k^{-5/3}$-decay throughout the
whole range of the wavenumbers.

In Figures~\ref{fig:p_cyclostrophic} and~\ref{fig:rho_cyclostrophic},
we show the contour plots of the pressure $\tilde p$ and density
$\tilde\rho$, respectively, of the cyclostrophic vortex, captured at
160, 170, 180 and 190 minutes of the elapsed model time. Note that
while the pressure assumes the form of a ``basin'' with closed level
curves (which is what naturally occurs in cyclones), the density
exhibits the same unrealistic variations as above in
Figure~\ref{fig:rho_inertial} for the inertial flow.

\section{Summary}
\label{sec:summary}

In the current work, we investigate the ability of our model of a
balanced compressible hard sphere gas flow \citep{Abr22} to produce
turbulent motions from a laminar initial condition in a purely
two-dimensional setting, which corresponds to the large scale dynamics
of the Earth atmosphere. The equations for such a flow are derived in
the same manner as in our previous work, with the additional condition
that a strong external acceleration in the downward direction,
together with an impenetrable bottom boundary of the domain, compress
the gas into a relatively thin horizontal slab.

We simulate two prototype scenarios of large scale atmospheric
dynamics -- an inertial jet, and a cyclostrophic vortex. In the
inertial jet scenario, we observe that an initially laminar straight
jet flow develops turbulent motions, bearing a close resemblance to
Reynolds' famous experiment \citep{Rey}. Moreover, the Fourier
spectrum of the kinetic energy of the turbulent part of the inertial
flow shows the Kolmogorov $k^{-5/3}$-decay at moderate scales, and the
$k^{-8/3}$-decay at small scales. In the cyclostrophic vortex
scenario, turbulent motions develop in the region of the maximal flow
speed, whereas the Kolmogorov $k^{-5/3}$-decay of the kinetic energy
spectrum is observed throughout the whole range of the Fourier
wavenumbers.

The main result of the current work is the surprising discovery that,
in our model of a balanced compressible hard sphere gas flow,
turbulent motions with Kolmogorov kinetic energy spectra develop not
only in a fully three-dimensional flow \citep{Abr22}, but also in a
restricted, two-dimensional setting. If our balanced gas flow
equations indeed constitute a faithful model of the actual phenomenon
of turbulence in gases, this result suggests that it should be
possible to capture large scale turbulent atmospheric features even in
simplified, lower-dimensional settings, such as those which are
typically used in long range climate change predictions.

\ack This work was supported by the Simons Foundation grant \#636144.

\renewcommand\bibfont

\end{document}